\documentclass[aps,physrev, twocolumn,superscriptaddress]{revtex4-2}

\usepackage[dvipsnames]{xcolor}
\usepackage{lipsum}
\usepackage{amsmath}
\usepackage[colorlinks=true, allcolors=blue]{hyperref}
\usepackage{graphicx}
\usepackage{babel}[english]
\usepackage{dcolumn}
\usepackage{bm}
\usepackage{float}
\usepackage{physics}
\usepackage{comment}

\begin{document}

\preprint{APS/123-QED}

\title{Epithelia Realize Nematopolar Topological Defect Structures}

\author{Tianxiang Ma}
\thanks{These authors contributed equally to this work}
\author{Niels de Graaf Sousa}
\thanks{These authors contributed equally to this work}
\author{Valeriia Grudtsyna}
\author{Farzan Vafa}
\email[]{farzan.vafa@nbi.ku.dk}
\author{Amin Doostmohammadi}
\email[]{doostmohammadi@nbi.ku.dk}
\affiliation{Niels Bohr Institute, University of Copenhagen, 2100 Blegdamsvej 17, Copenhagen, Denmark}

\date{\today}

\begin{abstract}

We introduce a shape-based polar order parameter that captures the structural asymmetry of cells within epithelial monolayers.
By combining bright-field imaging and traction force microscopy, we demonstrate that shape polarity serves as a unifying biomechanical metric, integrating the physical information encoded by nematic directors, principal stresses, and cellular motion.
Furthermore, we show that the tissue organizes into a mixed polar–nematic phase, characterized by the coexistence of integer ($\pm 1$) and half-integer ($\pm 1/2$) defects. 
Through mechanical perturbations, we demonstrate that both substrate stiffness and cell–cell adhesion modulate the density of these excitations and the length of domain walls binding like-signed positive half-integer defects.
Using a minimal continuum model of polar-nematic active matter, we establish that this mixed phase is fundamentally driven by the interplay of active stresses and polar-nematic elasticity.
These findings provide a direct experimental evidence that epithelial monolayers behave as nematopolar matter, in which coupled polar and nematic elastic interactions jointly shape the active state.

\end{abstract}

\maketitle




Topological defects are the fingerprints of broken continuous symmetry in matter. Recent studies increasingly identify topological defects in living materials, organizing the collective motion of systems ranging from cytoskeletal filaments to bacterial swarms~\cite{doostmohammadi2018active,martinez2019selection,dunkel2013fluid} and are attributed to cellular functions~\cite{saw2017topological,perez2019active,maroudas2021topological,guillamat2022integer}. The classification of topological defects is strictly dictated by the rotational symmetry of the underlying order parameter~\cite{shankar2022topological}. Nematic systems, characterized by their head-tail symmetry ($\mathbf{n} = -\mathbf{n}$), can minimize their elastic energy by forming half-integer charged disclinations ($s = \pm \tfrac{1}{2}$)~\cite{giomi2014defect}. In contrast, polar systems, whose order parameter is a vector field $\vec{p}$, minimize their elastic energy by forming integer topological charged defects ($s = \pm 1$)~\cite{juelicher2007active,ronning2023spontaneous}.


A striking feature of biological active matter is that constituent agents such as migrating cells or motile bacteria~\cite{han2025local,weiger2013real,saw2017topological}, are inherently polar, yet their collective dynamics are often described by active nematic formalisms, which produce half-integer defects and obscure the system's polar character~\cite{venkatesh2025interplay,giese2025polarity,bera2025cell}. 
Consequently, the fundamental interplay between individual particle polarity and global nematic order has remained experimentally inaccessible. Recent experimental work shows that several reconstituted and cellular systems are not constrained to a single symmetry imposed by the model; instead, polar and nematic interactions can coexist in a ``mixed'' phase~\cite{denk2020pattern,roostalu2018determinants,huber2018emergence}. However, coexistence of nematic and polar topological defects has not yet been observed.


In this Letter, we show that epithelial monolayers realize an active matter phase with mixed polar and nematic topological defects.
We introduce a shape-based cellular polarity, defined as a vectorial order parameter from the asymmetric geometry of each individual cell. 
This polarity order parameter recovers the conventional nematic texture associated with cell elongation, while retaining the head--tail information that is otherwise projected out by nematic descriptions. 

Notably, we find that the polarity order parameter in epithelial monolayers reveals a mixed topological state: its vector field exhibits integer polar defects, while the nematic moment of the same field exhibits extended-core structures whose endpoints are nematic half-integer defects. 
These extended cores form energetic domain walls binding pairs of like-signed half-integer defects, providing experimental evidence for the confinement mechanism predicted in nematopolar matter~\cite{vafa2025phase,mishra2025string,dinelli2026active,paik2026theory}, and reminiscent of quark confinement in high-energy physics~\cite{wilson1974confinement}. 
Crucially, these strings bind \emph{like-signed} defects, in contrast to earlier reports of string-like domain walls connecting \emph{oppositely charged} defect in two-dimensional classical XY models~\cite{Strings1985} and in sheared endothelial monolayers under externally imposed alignment~\cite{strings_amin_nat_phys}.
By combining experimental mechanical perturbations with a minimal continuum model, we show that activity drives both defect proliferation and core extension through the competition between polar and nematic elasticities.\\

\begin{figure}[ht]
\centering
\includegraphics[width=.99\linewidth]{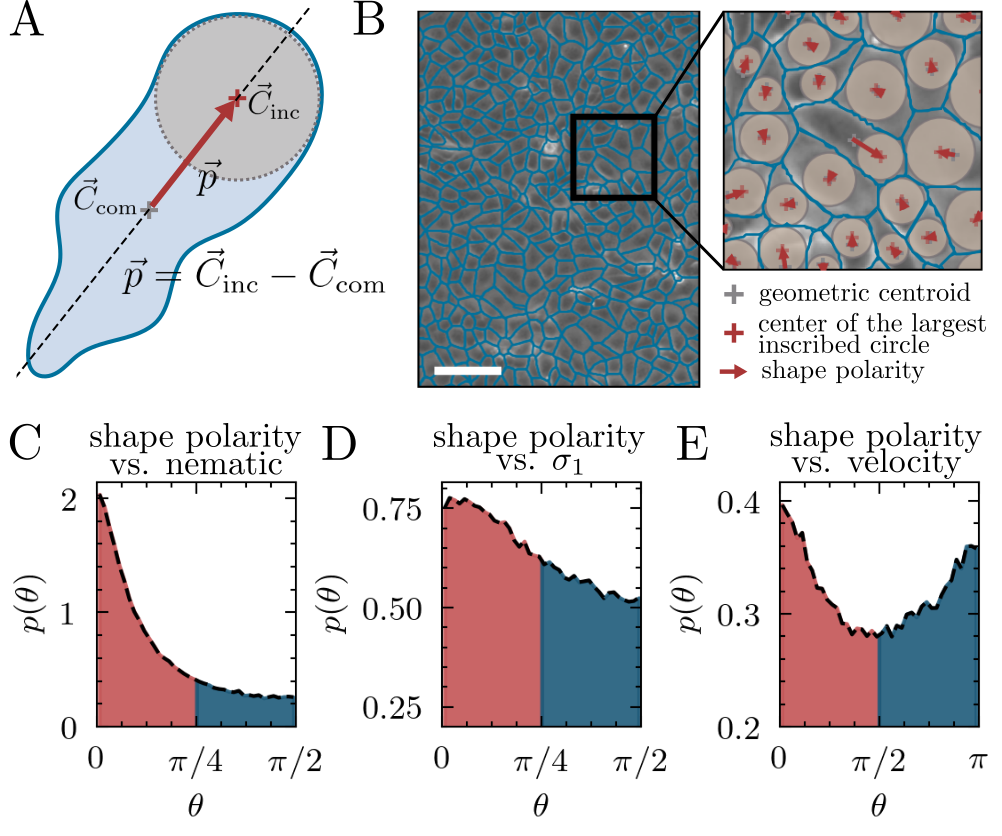}
\caption{\textbf{Shape polarity as a unifying structural order parameter.}
\textbf{(A)} Definition of the shape polarity vector 
$\vec{p} = \vec{C}_{\mathrm{inc}} - \vec{C}_{\mathrm{com}}$ from the geometric centroid 
$\vec{C}_{\mathrm{com}}$ to the center of the largest inscribed circle 
$\vec{C}_{\mathrm{inc}}$ of a single cell. 
\textbf{(B)} Example epithelial monolayer with segmented cell boundaries (blue), geometric centroids 
(gray crosses), centers of the largest inscribed circles (red crosses), and shape polarity vectors 
(red arrows). Scale bar: $100 \mu m$.
\textbf{(C)} Probability distribution $p(\theta)$ of the angle $\theta$ between shape polarity and the 
local nematic director reconstructed from cell elongation, showing strong alignment and thus that the 
symmetric moment of $\vec{p}$ reproduces the established nematic texture. 
\textbf{(D)} Distribution of $\theta$ between shape polarity and the maximal principal stress direction 
$\sigma_1$. 
\textbf{(E)} Distribution of $\theta$ between shape polarity and the cellular velocity, 
showing a bimodal form with minima near $\theta \simeq \pi/2$ and peaks at $0$ and $\pi$, indicating that 
$\vec{p}$ describes the principal axis of active motility. Shaded regions highlight the aligned (red) and non-aligned (blue) orientations.}
\label{fig:method}
\end{figure}

\noindent\textbf{Shape polarity as a unifying structural order parameter.--}
For each cell $i$, $\vec{p}_i$ is defined as the displacement vector from the geometric centroid $\vec{C}_{\mathrm{com}}$ to the center of the largest inscribed circle $\vec{C}_{\mathrm{inc}}$ (Fig.~\ref{fig:method}A, B). 
This definition captures the intrinsic front--rear asymmetry, \textit{i.e.}, the displacement of the bulk cell body relative to the leading edge. 

We demonstrate that this polar order parameter naturally encompasses the conventional nematic description of the tissue. 
By constructing the nematic tensor $\mathbf{Q} = \vec{p}\vec{p}^{\mathsf{T}} - \frac{p^2}{2}\mathbf{I}$, we recover the symmetric component of the order. 
Comparison of the director field $\mathbf{n}$ derived from this construction with the standard elongation axis reveals a dominant alignment tendency (Fig.~\ref{fig:method}C), which confirms that the symmetric moment of $\vec{p}$ faithfully reproduces the established nematic texture. In addition, $\vec{p}$ encodes key mechanobiological information consistent with active stress generation. By comparing $\mathbf{Q}$ with the intercellular principal stress axes measured from monolayer stress microscopy (see SM for methods~\cite{tambe2011collective,NIER20161625,anger2026tissue}), we observe a robust alignment (Fig.~\ref{fig:method}D).

Furthermore, unlike the headless elongation tensor, $\vec{p}$ constitutes an intrinsic vector order parameter. To validate its physical relevance, we compare $\vec{p}$ with the instantaneous cellular velocity. As shown in Fig.~\ref{fig:method}E, we observe an alignment characterized by bimodal distributions with deep minima at $\pi/2$ (perpendicular), and higher 0 to $\pi$ (align), indicating that $\vec{p}$ captures the principal axis of active motility. This bimodality reflects the local contact geometry: $\vec{p}$ preferentially points away from the nearest neighbor, whereas the velocity can point either away from or toward that neighbor (SM Fig.~S1), consistent with stochastic contact inhibition of locomotion~\cite{mayor2010keeping,kryvoruchko2026intrinsic}.

Consequently, $\vec{p}$ functions as the parent order parameter: its tensorial moment $\mathbf{Q}$ recovers the emergent nematic features, while its vector nature preserves the intrinsic polar order.\\

\begin{figure}[ht]
\centering
\includegraphics[width=.99\linewidth]{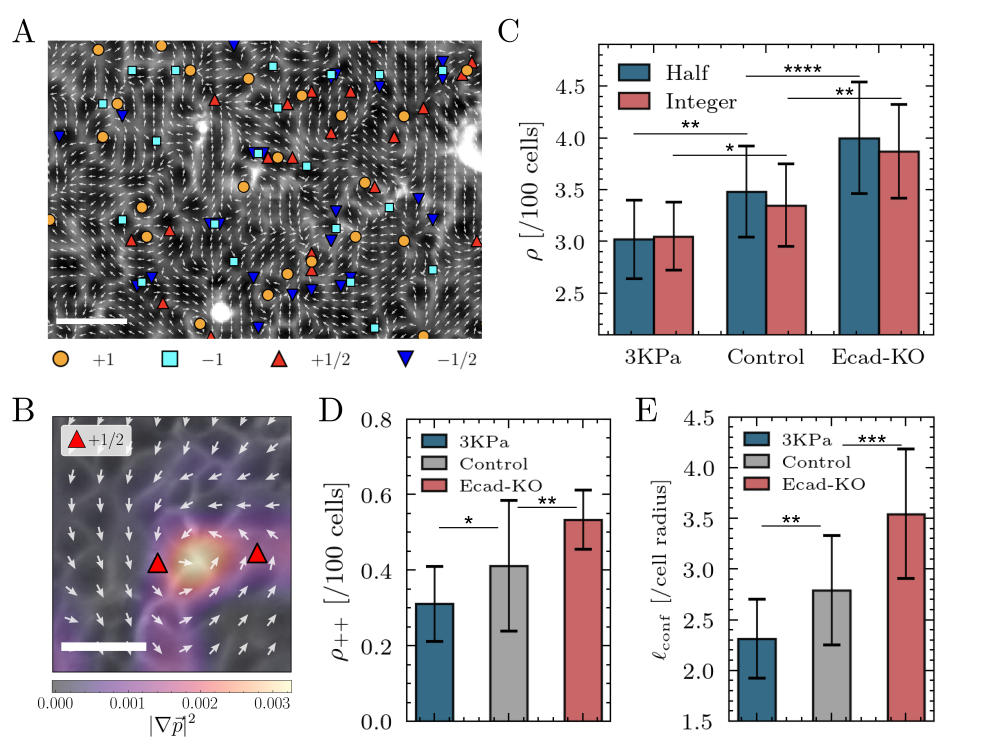}
    \caption{\textbf{Activity-dependent defect coexistence and topological confinement.}
    \textbf{(A)}~Snapshot of the shape polarity field overlaid with the bright-field image and the full- ($\pm 1$) as well as half-integer ($\pm 1/2$) topological defects. The system exhibits a mixed topology characterized by the coexistence of full-integer and half-integer defects. Scale bar: 100$\mu$m.
    \textbf{(B)}~Representative visualization of the polar elastic energy density overlaid with the shape polarity field and the bright-field image. High-energy ridges (bright regions) reveal positive strings that bind pairs of positive half-integer defects ($+1/2 \leftrightarrow +1/2$). Scale bar: 100$\mu$m.
    \textbf{(C)}~Variation of defect density with effective activity. The population of both half- and full-integer defects significantly increases monotonically across low (3 kPa), intermediate (Control), and high (E-cad KO) activity regimes.
    \textbf{(D, E)}~Statistics of confinement. Both the number of positive strings \textbf{(D)} and their length \textbf{(E)} increase with activity. Data are shown as mean $\pm$ SD (n = 15). Statistical significance was assessed using Welch’s \(t\)-tests; \(*p<0.05\), \(**p<0.01\), \(***p<0.001\), and \(****p<0.0001\).}
\label{fig: exp}
\end{figure}

\noindent\textbf{Defect coexistence and topological confinement.--}
To characterize the topological state of the monolayer, we calculate the local winding number of the shape-polarity field. 
In pure active nematics, the fundamental excitations are half-integer disclinations ($s=\pm 1/2$), whereas active polar fluids support only full-integer defects ($s=\pm 1$).
Interestingly, as shown in Fig.~\ref{fig: exp}A, the shape-polarity field exhibits a mixed topology, characterized by the stable coexistence of integer defects ($s=\pm1$) in the vector field $\vec{p}$ and half-integer defects ($s=\pm1/2$) in its tensorial moment $\mathbf{Q}$. 
This coexistence indicates that the system is not constrained by a single interaction but resides in a mixed regime where polar and nematic interactions compete.

This interaction conflict is physically reconciled through the introduction of positive strings that encapsulate +1 defects. These structures are characterized by an elongated nematic core, with endpoints that exhibit geometries resembling half-integer defects. While nematic relaxation proceeds via the annihilation of oppositely charged defects~\cite{giomi2013defect}, we observe the formation of persistent energetic domain walls binding pairs of positive half-integer defects ($+1/2 \leftrightarrow +1/2$). As visualized in Fig.~\ref{fig: exp}B, these strings appear as ridges of high polar elastic energy, representing regions of intense antiparallel alignment where the polar order is geometrically frustrated. By confining the polar discontinuity into a narrow domain wall, the system reconciles the local nematic charge with the global polar order. 

Having established the coexistence of polar and nematic topological structures, we next ask how this landscape is regulated by the two fundamental mechanical inputs underlying collective motion: cell--substrate and cell--cell interactions. 
To this end, we employ two experimental perturbations: substrate softening, which weakens cell--substrate traction and enhances cell--cell adhesion, and E-cadherin knockout (Ecad-KO), which produces the opposite shift. 
As quantified in Fig.~\ref{fig: exp}C, the densities of both half-integer and integer defects increase with stronger cell--substrate traction and decrease when cell--cell adhesion is enhanced. 
Crucially, the number of positive strings (Fig.~\ref{fig: exp}D) and their mean length (Fig.~\ref{fig: exp}E) follow the same trend: they are promoted by increased cell--substrate traction and suppressed by stronger cell--cell interactions. 
These results suggest that the mixed polar--nematic phase is an active topological state in which mechanical interactions modulate the defect nucleation and the extension of positive strings.\\

%
%

\noindent\textbf{Mechanism.--}
To explain the physical mechanism driving this topological confinement, we employ a minimal continuum model of polar active matter that explicitly accounts for the interplay between interactions by introducing a nematic elastic term~\cite{amiri2022unifying}. 
The system dynamics are governed by the minimization of a free energy functional $\mathcal{F}$ coupled to Stokes hydrodynamics driven by active stress $\sigma^a \sim -\zeta \mathbf{Q}$ (see End Matter for complete model). Crucially, the nematic tensor is not an independent field but is constructed directly from the polar order parameter as $\mathbf{Q} = \vec{p}\vec{p}^\mathsf{T} - \frac{p^2}{2}\mathbf{I}$, as in the experiments. This coupling is encoded in the free energy via two distinct elastic penalties:
\begin{equation}
\label{eq: first eq}
\mathcal{F} = \int d\mathbf{r} \left[
\frac{A_p}{2}\left(1-\vec{p}^{\,2} \right)^2
+ \frac{K_p}{2}(\nabla \vec{p})^2
+ \frac{K_n}{2}(\nabla \mathbf{Q})^2
\right].
\end{equation}
Here, the polar elasticity $K_p$ penalizes gradients in the vector field, while the nematic elasticity $K_n$ penalizes gradients in the dependent tensor field. The parameter $\zeta$ controls the energy injection rate, serving as the theoretical analogue to our experimental modulation.

\begin{figure}[t]
    \centering
    \includegraphics[width=1\linewidth]{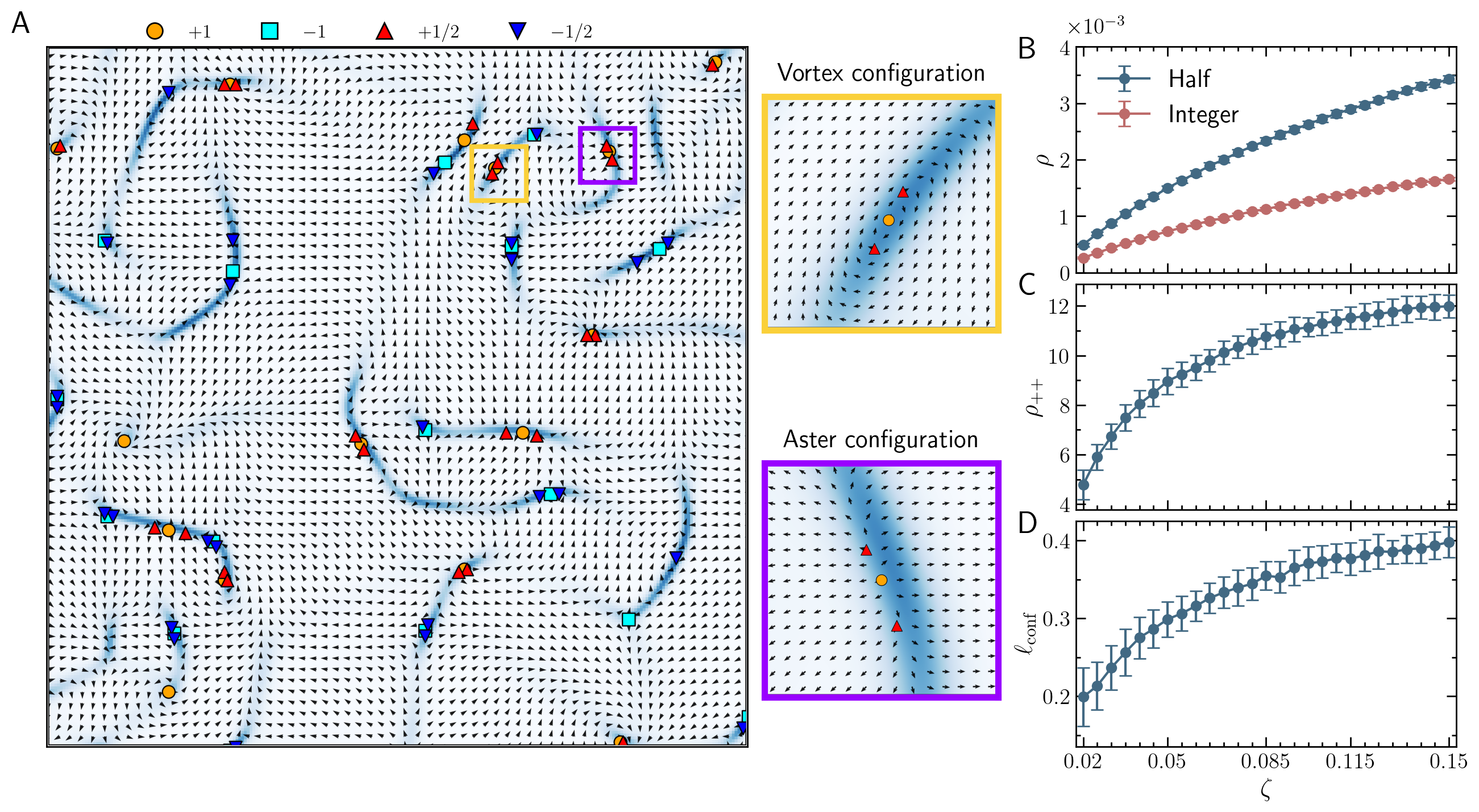}
    \caption{\textbf{Defect coexistence and activity-driven strings in the continuum simulations.} 
    \textbf{(A)} Snapshot of the polarization field displaying the coexistence of mixed topological defects. Colorbar represents the elastic energy density.
    \textbf{(B)} Defect density as a function of the activity parameter $\zeta$ for both integer and half-integer topological charges, exhibiting a monotonic increase with increasing activity.
    \textbf{(C)} Density of positive strings as a function of activity. The string density is normalized by the square root of the $+1/2$ defect number.
    \textbf{(D)} Mean positive string length as a function of the activity parameter. The length is normalized by the characteristic defect spacing, $\sqrt{L^{2}/N_{+1/2}}$.
    Simulations use elastic constants $K_n = K_p = 0.5$.}
    \label{fig: sim}
\end{figure}

Simulating this model in the ``mixed symmetry'' regime ($K_p\approx K_n$) successfully reproduces the experimental phenomenology, finding regimes of mixed topological charge (Fig.~\ref{fig: sim}A). By increasing the active stress $\zeta$ while holding polar and nematic elasticities
constant, the system transitions from a quiescent state to a chaotic state populated by both half and full-integer defects. We observe that the density of integer defects and the number of positive strings scale monotonically with $\zeta$ (Fig.~\ref{fig: sim}B and C).

Positive strings emerge as a consequence of the nematic elastic interactions, $K_n$. In a purely polar system, antiparallel alignment would incur a prohibitive energetic cost, precluding the formation of such domain walls. While the polar elasticity $K_p$ confines the two $+1/2$ defects to form a +1 integer defect, the nematic elasticity $K_n$ allows the defects to extend, forming a nematically aligned region within the defect core, resembling a string morphology. The energetic cost associated with the formation of a string is given by~\cite{vafa2025phase,selinger2024introduction},
\begin{equation}
    \label{eq: energy of forming a string}
    \mathcal{F_{\mathrm{string}}}= \frac{\pi}{4} K_n \ln{\frac{L}{\ell_{s}}}+ T\ell_{s},
\end{equation}
where $L$ indicates the system size and $\ell_{s}$ the length of the string. The first term accounts for the Coulomb repulsion between both defects~\cite{shankar2018defect}, while $T$ indicates the string tension or domain wall energy which takes the following form,
\begin{equation}
    T = \frac{A_p}{2}\sqrt{\frac{K_p/2}{A_p/2}} = \frac{1}{2}\sqrt{K_p A_p}.
\end{equation}

It is worth emphasising that this confinement mechanism differs from recently reported multicellular domain walls in sheared endothelial monolayers, where externally imposed alignment connects \emph{oppositely charged} nematic defects~\cite{strings_amin_nat_phys}. Here, by contrast, the strings bind \emph{like-signed} positive half-integer defects as a direct consequence of nematopolar interactions.

The dynamics of the positive string are governed by the competition between a passive restoring force and an active motile force. The passive contribution follows from the string free energy as
\begin{equation}
    \label{eq: passive force}
    \mathbf{F}_P = -\frac{\partial F_{\mathrm{string}}}{\partial \ell_s} = \frac{\pi}{4}\frac{K_n}{\ell_s} - T.
\end{equation}
Activity introduces an additional motile force $\mathbf{F}_M = \zeta\nabla\cdot \mathbf{Q}$. Since this force vanishes for a $-1/2$ defect but is non-zero for a $+1/2$ defect, only the positive string length receives a motile contribution. The motile force is directed along the defect polarization axis, which forms an angle $\psi$ with the string: $\psi = 0$ corresponds to an aster, $\psi = \pi/2$ to a vortex, and intermediate values to a spiral configuration. The component of the motile force parallel to the string is 
\begin{equation}
    \mathbf{F}_M = \cos(2\psi) \frac{\pi}{4} \frac{\zeta}{a}, 
\end{equation}
where $a$ is the defect core radius~\cite{vafa2020multi-defect,vafa2024periodic}. Because $\mathbf{F}_M$ is independent of the string length, activity renormalizes the string tension, yielding an effective tension of the form
\begin{equation}
    T_{\mathrm{eff}} = \frac{1}{2}\sqrt{K_p A_p} + \cos(2\psi) \frac{\pi}{4}\frac{\zeta}{a}.
\end{equation}
Balancing the passive restoring force against the active motile force (Eq. \ref{eq: passive force}) yields the equilibrium string length,
\begin{equation}
\ell_{s} = \frac{K_n}{\frac{2}{\pi}\sqrt{K_p A_p} + \cos(2\psi) \frac{\zeta}{a}}.
\label{eq: minimized_string_length}
\end{equation}

This analytical relation captures the activity‑induced increase in string length observed in simulations (Fig.~\ref{fig: sim}D), and provides a direct interpretation of the experimental trend in Fig.~\ref{fig: exp}E.
Traction force microscopy shows that substrate softening (3~kPa) suppresses collective motility and cell--substrate traction, whereas E-cadherin knockout (Ecad-KO) enhances both (End Matter Fig.~\ref{fig:traction_vel}).
Recent work has shown that, in active cellular monolayers, the active stress increases with the mean traction force~\cite{han2025local}. 
Therefore, our traction force measurements place the three perturbations in an increasing activity sequence from 3~kPa to control and then to Ecad-KO.
The mean positive string length follows this same increase   (Fig.~\ref{fig: exp}E), supporting an activity-controlled mechanism for string extension.

Beyond setting the overall scale of string extension, Eq.~\ref{eq: minimized_string_length} also reveals a configuration-dependent response to activity. In the extensile regime, aster configurations become more tightly bound while vortex configurations loosen, whereas in the contractile regime the opposite trend occurs. Simulations display both configurations; however, Eq.~\ref{eq: minimized_string_length} shows that the defect string whose length increases with activity, namely vortices in extensile systems and asters in contractile systems, elongates more than the one that shortens, leading to an overall increase of the total string length with activity (see End Matter for details). These results demonstrate that activity drives the system further from equilibrium by enhancing the spatial separation of bounded $+1/2$ defects and promoting the expansion of aligned nematic domains. As shown in Eq.~\ref{eq: minimized_string_length}, the purely polar limit ($K_n = 0$) precludes the dissociation of $+1$ defects; since nematic interactions are suppressed, the energetic pathway for defect unbinding is unavailable. Conversely, in the purely equilibrium nematic limit ($K_p = 0$ and $\zeta=0$), the string length diverges. The $+1$ defects consequently dissociate into unbounded $+1/2$ pairs, recovering the characteristic dynamics of active nematics. Contrary to active polar gels~\cite{kruse2004asters,kruse2005generic} experiments and simulations reveal the coexistence of all types of $+1$ defect configurations (Fig.~\ref{fig: exp}A and Fig.~\ref{fig: sim}A): aster, spirals and vortices characterized by the polarization orientation or $\psi$ ~\cite{elgeti2011defect}. This coexistence arises because the system operates in the flow-tumbling regime, where no defect configuration is stationary.\\


\noindent\textbf{Discussion.--}
Even though individual cells are polar objects, tissue-scale dynamics are often interpreted through nematic order parameters derived from cell elongation.
Here, by defining a shape-based polarity for each cell, we show that the nematic description is a projection of a richer polar structure. 
The vectorial polarity retains head--tail information, while its tensorial moment recovers the conventional nematic texture.
This establishes epithelial monolayers as an active system that is not constrained by a single broken symmetry but operates in a crossover regime governed by the competition between polar and nematic elasticities.


Within this mixed polar--nematic regime, our experiments reveal topological confinement in a biological active fluid, manifested as strings connecting positive half-integer defects.
We identify the string structures connecting structures that resemble positive half-integer defects ($+1/2 \leftrightarrow +1/2$) not as transient fluctuations, but as energetic domain walls---physical manifestations of the ``branch cuts'' required to reconcile nematic topological charges with a polar order parameter. 
Together, these observations provide experimental evidence for the confinement mechanism predicted in nematopolar matter~\cite{vafa2025phase,mishra2025string,dinelli2026active,paik2026theory}. 
A detailed analysis of the mixed-symmetry model reveals that confinement emerges as a dynamical state, governed by the competitive interplay between nematic and polar elastic interactions.
To validate the proposed model, we examine the dependence of string density and length on the activity parameter $\zeta$. From an analytical perspective, we demonstrate that the global string length increases monotonically with activity confirming the behaviour seen in experiments and continuum simulations. Together, these results connect polar flocking and active nematic turbulence within a single framework.


More broadly, our findings establish nematopolar active matter as a potential framework for understanding tissue organization in systems where polarity coexists with orientational order. 
This is particularly relevant for multicellular processes in which chemical, adhesive, and mechanical cues polarize individual cells while coordinated interactions organize collective motion, including morphogenesis, wound repair, regeneration, and cancer invasion~\cite{friedl2009collective,scarpa2024perspectives,cheung2025collective,peglion2023polarity}. 
By retaining cell polarity while recovering the conventional nematic texture, our approach reveals mechanobiological and topological information otherwise projected out by purely nematic descriptions. 
It therefore provides a route to identify and characterize mixed polar--nematic organization in multicellular systems, with potential relevance to diverse forms of tissue organization in health and disease.
~\\

\begin{acknowledgments}
F.V. and A. D. acknowledge funding from the Novo Nordisk Foundation (grant No. NNF18SA0035142). A.D. additionally acknowledges funding from the Novo Nordisk Foundation NERD programme (grant No. NNF21OC0068687), Villum Fonden (Grant no. 29476), and the European Union (ERC, PhysCoMeT, 101041418). Views and opinions expressed are however those of the authors only and do not necessarily reflect those of the European Union or the European Research Council. Neither the European Union nor the granting authority can be held responsible for them. The Tycho supercomputer hosted at the SCIENCE HPC center at the University of Copenhagen was used for supporting this work.
\end{acknowledgments}


\nocite{*}

\bibliography{apssamp}
\appendix

\section{Model and Simulation Details}

We employ a general framework to describe the dynamics of the polarity field $\vec{p}(\mathbf{r},t)$ of particles in an incompressible fluid with velocity field $\vec{v}(r,t)$ and $\nabla \cdot \vec{v}=0$. The equation for polarization is given by \cite{juelicher2007active},
\begin{equation}
    \label{eq: sm pol}
    \partial_t \vec{p} + \vec{v}\cdot \nabla \vec{p} + \bm{\omega} \cdot \vec{p} = \frac{1}{\gamma}\vec{h}-\lambda \mathbf{E} \cdot \vec{p},
\end{equation}
where $\gamma$ is the rotational viscosity that controls relaxation of
the polarity field to the minimum of the free energy through
the molecular field $\vec{h}=-\frac{\delta\mathcal{F}[\vec{p}]}{\delta \vec{p}}$. Here $\mathbf{E}$ and $\bm{\omega}$ are the strain and vorticity tensors, respectively.

The dynamics of the velocity are governed by the Navier-Stokes equation,
\begin{equation}
    \label{eq: sm vel}
    \rho \,( \partial_t \vec{v} + \vec{v} \cdot \nabla \vec{v})=\nabla \cdot \bm{\sigma},
\end{equation}
where $\rho$ is the fluid density. The stress $\bm{\sigma}$ has three components; (i) viscous $\bm{\sigma}^{vis}=2 \eta \mathbf{E}$, (ii) passive $\bm{\sigma}^{pas}=-P\mathbf{I} + \frac{\lambda +1}{2} \vec{p}\vec{h} + \frac{\lambda-1}{2}\vec{h}\vec{p} - \frac{\lambda}{2}( \vec{p}\cdot \vec{h}) \mathbf{I}$ and (iii) active $\bm{\sigma}^{act}=- \zeta (\vec{p}\vec{p}^\mathsf{T} - \frac{p^2}{2}\mathbf{I})$. Here $\eta$ is the viscosity constant, $P$ is the isotropic pressure, $\lambda$ is the aligning parameter and $\zeta$ the activity coefficient. 

We numerically integrate Eq. \ref{eq: sm pol} and Eq. \ref{eq: sm vel}, using a hybrid Lattice-Boltzmann algorithm, with Periodic-Boundary-Conditions (PBC) and with simulation parameters set to the following values unless otherwise specified: $L=1024$, $\rho=40$, $\gamma=10$, $\lambda=0.1$, $K_p=0.5$, $K_n=0.5$, $A_p=0.1$ and $\eta=20$.

\section{Overall String length respect activity}

In this section, we show that the mean string length increases monotonically with activity. The string length associated with a defect of orientation $\psi$ is given by
\begin{equation}
    \ell_s(\psi) = \frac{aK_n}{\frac{2a}{\pi}\sqrt{K_p A_p} + \zeta\cos(2\psi)}.
\end{equation}
Since a reversal of the activity sign interchanges defect orientations, the mean string length depends only on $|\zeta|$. Averaging over all orientations $\psi \in [0, \pi/2]$, we obtain
\begin{equation}
    \bar{\ell}_s = \frac{2}{\pi}\int_0^{\pi/2} \ell_s(\psi)\,d\psi = \frac{aK_n}{\sqrt{\left(\frac{2a}{\pi}\right)^2 K_p A_p - \zeta^2}},
\end{equation}
which is strictly increasing in $|\zeta|$, consistent with the global growth of string length observed in both simulations and experiments. For small activity, writing $A = \frac{2a}{\pi}\sqrt{K_p A_p}$, we expand to leading order in $\zeta$,
\begin{equation}
    \bar{\ell}_s = \frac{aK_n}{A}\left(1 - \frac{\zeta^2}{A^2}\right)^{-\frac{1}{2}} \approx \frac{\pi K_n}{2\sqrt{K_p A_p}}\left(1 + \frac{\pi^2 \zeta^2}{8 a^2 K_p A_p} \right),
\end{equation}
where the correction grows as $\zeta^2$, consistent with the symmetry $\bar{\ell}_s(|\zeta|)$. At $\zeta=0$, this correctly recovers the equilibrium string length $\bar{\ell}_s|_{\zeta=0} = \frac{\pi K_n}{2\sqrt{K_p A_p}}$.

\section{Mechanical perturbations modulate multicellular activity}

\begin{figure}[h!]
    \centering
    \includegraphics[width=.99\linewidth]{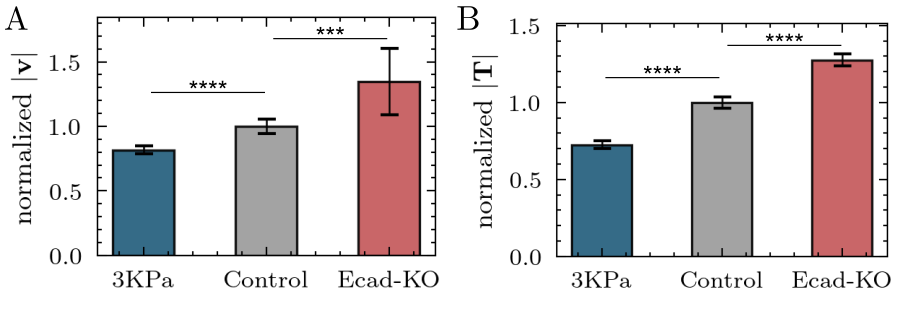}
    \caption{\textbf{Mechanical perturbations tune multicellular activity.}
    \textbf{(A)} Normalized mean cellular speed, $|\mathbf{v}|$, under substrate and adhesion perturbations. Substrate softening decreases collective motility, whereas E-cadherin knockout increases it.
    \textbf{(B)} Normalized mean traction magnitude, $|\mathbf{T}|$, measured by traction force microscopy. Relative to control, 3~kPa substrates reduce cell--substrate traction, whereas Ecad-KO increases it. Data are shown as mean $\pm$ SD (n = 15). Statistical significance was assessed using Welch’s \(t\)-tests; \(***p<0.001\), and \(****p<0.0001\).}
    \label{fig:traction_vel}
\end{figure}

As shown in Fig.~\ref{fig:traction_vel}, both collective motility and cell--substrate traction are strongly modulated by cell--substrate and cell--cell interactions.
Compared to wild-type cells on 15~kPa substrates (control), cells on softer 3~kPa substrates (3~kPa) exhibit reduced mean velocity and reduced traction magnitude.
Conversely, E-cadherin knockout cells on 15~kPa substrates (Ecad-KO) show enhanced mean velocity and traction. 
Together, these suggest the 3~kPa, control, and Ecad-KO conditions as low-, intermediate-, and high-activity states, respectively.

\end{document}